\documentclass[12pt,letterpaper]{article}
\usepackage{graphicx,amsmath,amssymb,psfrag,mathabx,bbm,tikz,url} %hyperref

\newcommand{\matr}[1]{\mathbf{#1}}
\newcommand{\vect}[1]{\mathbf{#1}}

\newcommand{\code}{\mathcal{C}}

\newcommand{\matG}{\matr{G}}

\newcommand{\vecGamma}{\mbox{\boldmath$\gamma$} }

\newcommand{\vecWs}{\vect{w}_\vect{s}}
\newcommand{\vecCs}{\vect{c}_\vect{s}}
\newcommand{\factorGraph}{\mathcal{G}}

\newcommand{\bia}[1]{B_{i\rightarrow a}^{(#1)}}

\newcommand{\bsaa}[1]{B_{\vect{s}_a\rightarrow a}^{(#1)}}
\newcommand{\sai}[1]{S_{a\rightarrow i}^{(#1)}}
\newcommand{\sbi}[1]{S_{b\rightarrow i}^{(#1)}}

\newcommand{\checks}{C}
\newcommand{\infobits}{V}
\newcommand{\allbits}{\overline{V}}

\newcommand{\dprod}[2]{\langle #1 | #2 \rangle}

\newcommand{\ddens}[1]{#1} %\mathrm{#1}

\newcommand{\convcond}{\mathfrak{C}}

\newcommand{\wsou}{w_{sou}}
\newcommand{\winfo}{w_{info}}

\newcommand{\pnummin}{\texttt{num\_min}}
\newcommand{\pthreshold}{\texttt{t}}
\newcommand{\pnummax}{\texttt{num\_max}}
\newcommand{\pstartdamp}{\texttt{start\_damp}}
\newcommand{\pmaxiter}{\texttt{max\_iter}}

\begin{document}
\renewcommand{\textfraction}{0}
\title{Binary quantization using Belief Propagation with decimation over factor graphs of LDGM codes\footnote{This research was supported by Air Force Research Laboratory, Air Force Material Command, USAF, under the research grant FA8750-04-1-0112.
%The U.S. Government is authorized to reproduce and distribute reprints for Governmental purposes notwithstanding any copyright notation there on. The views and conclusions contained herein are those of the authors and should not be interpreted as necessarily representing the official policies, either expressed or implied, of Air Force Research Laboratory, or the U.S. Government.
}}

\author{\normalsize Tom\'{a}\v{s} Filler, Jessica Fridrich \\ \small Dept. of Electrical and Computer Engineering\\[-5pt]\small SUNY Binghamton, Binghamton, NY 13902-6000, USA\\[-5pt] \small \{tomas.filler,fridrich\}@binghamton.edu}
\date{}
\maketitle
\thispagestyle{empty}
\begin{abstract}
We propose a new algorithm for binary quantization based on the Belief Propagation algorithm with decimation over factor graphs of Low Density Generator Matrix (LDGM) codes. This algorithm, which we call Bias Propagation (BiP), can be considered as a special case of the Survey Propagation algorithm proposed for binary quantization by Wainwright et al. \cite{Wai05}. It achieves the same near-optimal rate-distortion performance with a substantially simpler framework and 10--100 times faster implementation. We thus challenge the widespread belief that binary quantization based on sparse linear codes cannot be solved by simple Belief Propagation algorithms. Finally, we give examples of suitably irregular LDGM codes that work with the BiP algorithm and show their performance.
%An important advantage of our reformulation using Belief Propagation is that the algorithm can be analyzed using standard tools previously developed for Density Evolution. We derive a necessary condition that the node degree distributions of the associated factor graphs must satisfy to obtain good BiP quantizers.
\end{abstract}
\normalsize

\section{Introduction}
Binary quantization is an important problem for lossy source coding and other fields, such as information hiding \cite{Fri07Spie}. The recent work of Wainwright et al. \cite{Wai05} shows that Low Density Generator Matrix (LDGM) codes combined with Survey Propagation (SP) message-passing algorithms can be used to achieve near-optimal binary quantization in practice. Theoretical properties of regular LDGM codes for binary quantization were studied by Martinian et al. \cite{Mar06}. These results motivated us to study practical algorithms for binary quantization using LDGM codes.

% Recently, Ciliberti et al. \cite{Cil05} described an approach for binary quantization based on general random gates and SP algorithm. They had to depart from ordinary parity check gates (constraint satisfaction problem based on binary linear codes), because the approach based on SP algorithm did not give satisfactory results. The authors constrained themselves solely to regular codes (all check nodes were connected to the same number of bits). 
% In fact, BP is much simpler algorithm than SP, thus the implementation of the whole compression scheme is much faster in practice.

% JF: Mne se zda, ze ta prace Cilibertiho et al. je tady out of place. Co presne tou citaci chces rict? Ze SP + regular codes do not work? Neni to pro nas mimo misu?
% TF: dobra, priznavam ze je to trosku mimo :)

% TF: opravil jsem v nasledujici radce can not na cannot. Nebo to tu bylo tak zamerne?
In this paper, we challenge the claim that lossy source coding based on pure Belief Propagation (BP) as opposed to SP cannot give satisfactory results. We propose an approach based on pure Belief Propagation for quantizing random Bernoulli source with $p=\frac{1}{2}$. This algorithm, which we call Bias Propagation (BiP), achieves the same near-optimal rate-distortion performance in comparison with \cite{Wai05}. This work has been motivated by applications of binary quantization in steganography (information hiding), where the problem appears in a slightly more general setting called \emph{weighted binary quantization}.

In particular, we are quantizing a random $n$-bit source sequence $\vect{s}$ to the nearest codeword $\vect{c}_\vect{s}$ from an LDGM code $\code$ with rate $R=\frac{m}{n}$. In the weighted binary quantization problem, we replace the Hamming distance with the weighted distortion measure $d_\varrho(\vect{s},\vect{c}_\vect{s})=\frac{1}{n}\sum_{i=1}^n\varrho_i|\vect{s}_i-(\vect{c}_\vect{s})_i|$, where $\varrho_i\in[0,1]$. We call the vector $\varrho=(\varrho_1,\ldots,\varrho_n)$ the \emph{profile} of the weighted binary quantization problem. Our goal is to minimize the average distortion $D_\varrho=\mathbb{E}[d_\varrho(\vect{s},\vect{c}_\vect{s})]$ for a given profile $\varrho$, where $\mathbb{E}[\cdot]$ is the expectation taken over all possible source sequences $\vect{s}$.  In case of \emph{uniform profile} $\varrho=(1,\ldots,1)$, we denote the average distortion as $D$. The rate-distortion function is in the form $R(D)=1-H(D)$ for $D\in[0,0.5]$ and $0$ otherwise, where $H$ is the binary entropy function. Rate-distortion functions for other profiles can be found in \cite{Fri07Spie}.

The rest of this paper is structured as follows. In Section 2, we describe LDGM codes and introduce notation. In Section 3, we present a complete derivation of the BiP algorithm using Belief Propagation over factor graphs of LDGM codes. In Section 4, we briefly discuss the reason why an approach based on the BP algorithm should give us satisfactory results. We make a connection to the recent work of Murayama \cite{Mur04}. The paper is concluded in Section 5, where we present some experimental results.

\section{LDGM code representation}
\begin{figure}[t]
	\begin{center}

	\def\source#1:#2:#3;{
		\draw (#1) node {\footnotesize #2} (#1)--+(0,0.7)--+(-0.15,0.7)--+(+0.15,0.7) node[pos=0.5, above] {$#3$};
		\draw[color=black] (#1) +(-0.175,0.7) rectangle +(+0.175,0.7);
	}

	{ \begin{minipage}{3.3cm}
		$\mathbf{G}=\left(\begin{array}{ccc}
		1 & 0 & 0\\
		0 & 1 & 0\\
		1 & 0 & 1\\
		1 & 1 & 1\\
		0 & 0 & 1\\
		0 & 0 & 1\end{array}\right)$
	\end{minipage} } \hspace*{0.5cm}
	{ \begin{minipage}{5.5cm}
		{\begin{tikzpicture}[scale=1]
			\tikzstyle{info_bit}=[circle, draw, minimum size=4mm]
			\tikzstyle{check}=[rectangle, draw, minimum height=4.5mm, minimum width=4.5mm]
			\draw (0,0) node[check] (f1) {}
				(1,0) node[check] (f2) {}
				(2,0) node[check] (f3) {}
				(3,0) node[check] (f4) {}
				(4,0) node[check] (f5) {}
				(5,0) node[check] (f6) {};
			\draw (1.5,-1.25) node[info_bit] (w1) {}
				(2.5,-1.25) node[info_bit] (w2) {}
				(3.5,-1.25) node[info_bit] (w3) {};
			\source f1:{\scriptsize a}:\mathbf{s}_a;
			\source f2:{\scriptsize b}:\mathbf{s}_b;
			\source f3:{\scriptsize c}:\mathbf{s}_c;
			\source f4:{\scriptsize d}:\mathbf{s}_d;
			\source f5:{\scriptsize e}:\mathbf{s}_e;
			\source f6:{\scriptsize f}:\mathbf{s}_f;
		
			\draw (w1)--(f1); \draw (w1)--(f3); \draw (w1)--(f4);
			\draw (w2)--(f2); \draw (w2)--(f4);
			\draw (w3)--(f3); \draw (w3)--(f5); \draw (w3)--(f4); \draw (w3)--(f6);
		
			\draw (w1) node[below=8pt] {$\mathbf{w}_i$};
			\draw (w2) node[below=8pt] {$\mathbf{w}_j$};
			\draw (w3) node[below=8pt] {$\mathbf{w}_k$};
			%\draw (7.28,0.8) node {$\vect{s}_f$ ... source bit};
			%\draw (7.9,0) node {$f$  ... check node with};
			%\draw (7.7,-0.4) node {degree 1};
			%\draw (7,-1.4) node {$\vect{w}_k$ ... info bit with degree 4};
		\end{tikzpicture} \vspace*{-0.6cm} }
	\end{minipage} }

	\end{center}\vspace*{-0.3cm}

	\caption{Factor graph representation of a linear code with generator matrix $\mathbf{G}$.}
	\label{fig:factor-graph}
\end{figure}
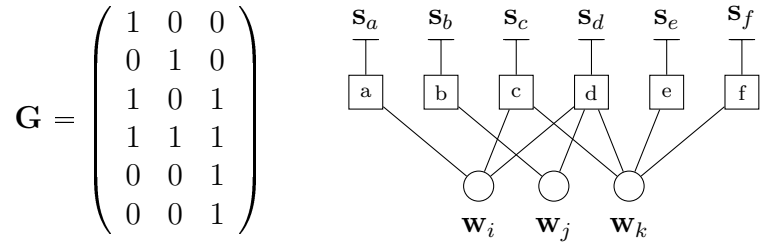

Codes based on sparse generator matrices are duals of LDPC codes. For a given linear code $\code$ with generator matrix $\matG\in\{0,1\}^{n\times m}$, we define the factor graph of this code as a graph $\mathcal{G}=(V, C, E)$ with $n$ \emph{check nodes} $C=\{1,\ldots,n\}$, $m$ \emph{information bits} $V=\{1,\ldots,m\}$, and $n$ \emph{source bits}. An example of a factor graph can be seen in Figure~\ref{fig:factor-graph}. We will use variables $a,b,c\in\checks$ to denote the check nodes and variables $i,j,k\in\infobits$ to denote the information bits. Each check node $a\in\checks$ has its associated source bit $\vect{s}_a$. Check node $a$ is connected with information bit $i$ , $(a,i)\in E$, if $\matG_{a,i}=1$. Finally, we define the sets $C(i) = \{a\in\checks\,|\,(a,i)\in E\}$, $V(a) = \{i\in\infobits\,|\,(a,i)\in E\}$, and $\allbits(a) = \infobits(a)\cup\{\vect{s}_a\}$.

The factor graph of an LDGM code is obtained randomly using degree distributions from the edge perspective $(\rho,\lambda)$, $\rho(x) = \sum_{i=1}^{d_R} \rho_i x^{i-1}$ and $\lambda(x) = \sum_{i=1}^{d_L} \lambda_i x^{i-1}$, where $\rho_i$ and $\lambda_i$ denote the portion of all edges connected to check nodes and information bits with degree $i$, respectively. The degree of the check node $a$ is defined as the number of connected information bits (we do not count the associated source bit).

\section{Bias Propagation algorithm}
Let $\code$ be an LDGM code with generator matrix $\matG\in\{0,1\}^{n\times m}$, $\vect{s}\in\{0,1\}^n$ a fixed source sequence, and $\varrho$ a given profile. For a given constant vector $\vecGamma=(\gamma_1,\ldots,\gamma_n)$, we define the following conditional probability distribution over LDGM codewords
\begin{align}
P(\vect{w}|\vect{s}; \vecGamma) =
\frac{1}{Z} e^{-2 \dprod{\vecGamma}{\matG\vect{w}-\vect{s}} }, \label{eq:bip-probability-distribution}
\end{align}
where $\dprod{\vecGamma}{\matG\vect{w}-\vect{s}}$ is the dot product of vectors $\vecGamma$ and $\matG\vect{w}-\vect{s}$ (calculated in binary arithmetic),  $\vect{w}\in\{0,1\}^m$, and $Z$ is a normalization constant $Z = \sum_{\vect{c}\in\code} e^{-2\dprod{\vecGamma}{\vect{c}-\vect{s}}}$. The most probable codeword $\vecCs=\matG\vecWs$ is the optimal solution to our original problem. The vector $\vecGamma$ should be determined from the profile $\varrho$.

\begin{figure}[t]
	\begin{center}
	{Bias Propagation Algorithm (BiP)}\\
	\fbox{
	\begin{minipage}{15.5cm}
		(a) \underline{pseudo-code}\\
		\includegraphics*[viewport=70 630 580 780, scale=0.87]{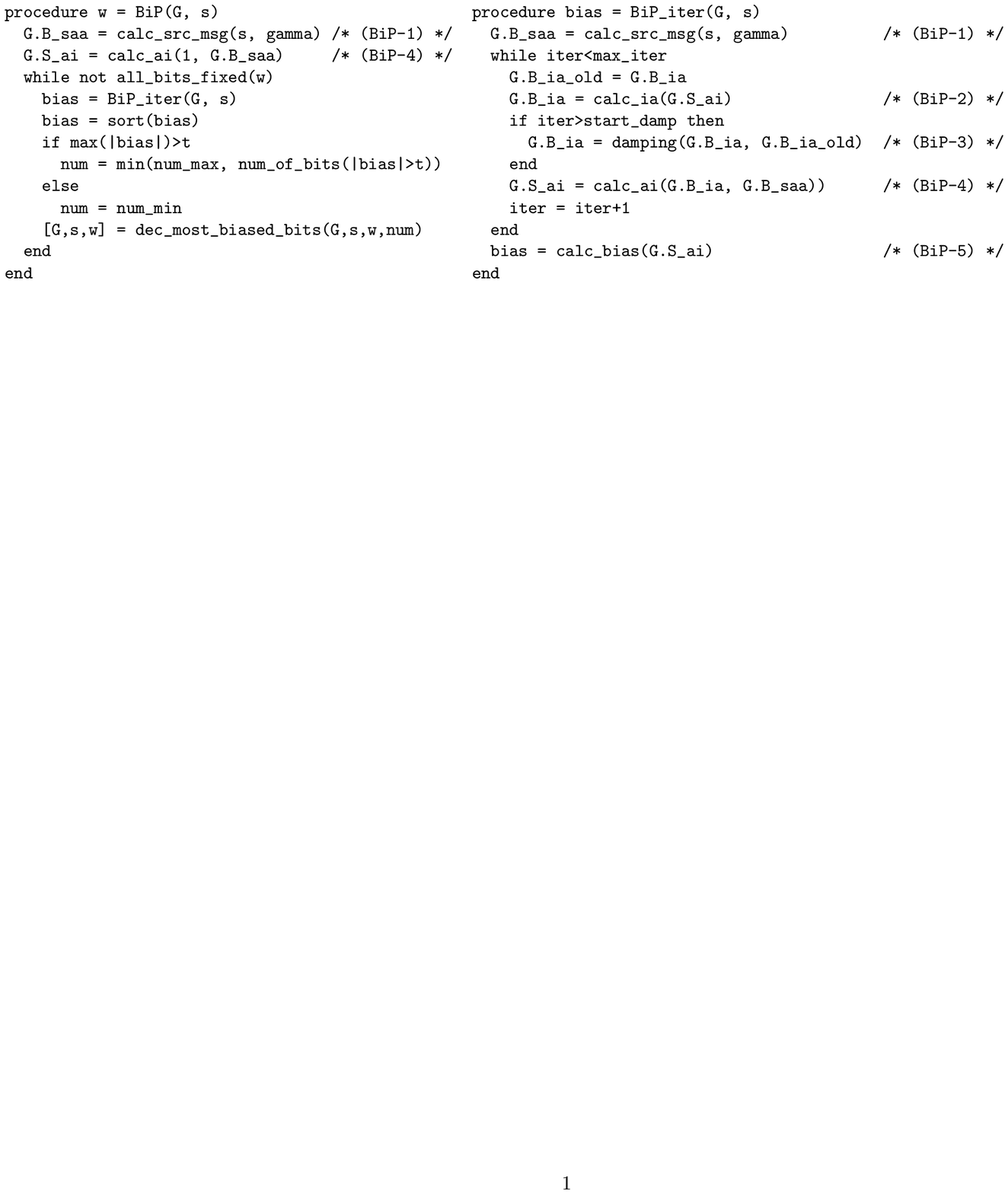}\\
		(b) \underline{message-passing update rules}\\
		\includegraphics*[viewport=0 500 620 780, scale=0.77]{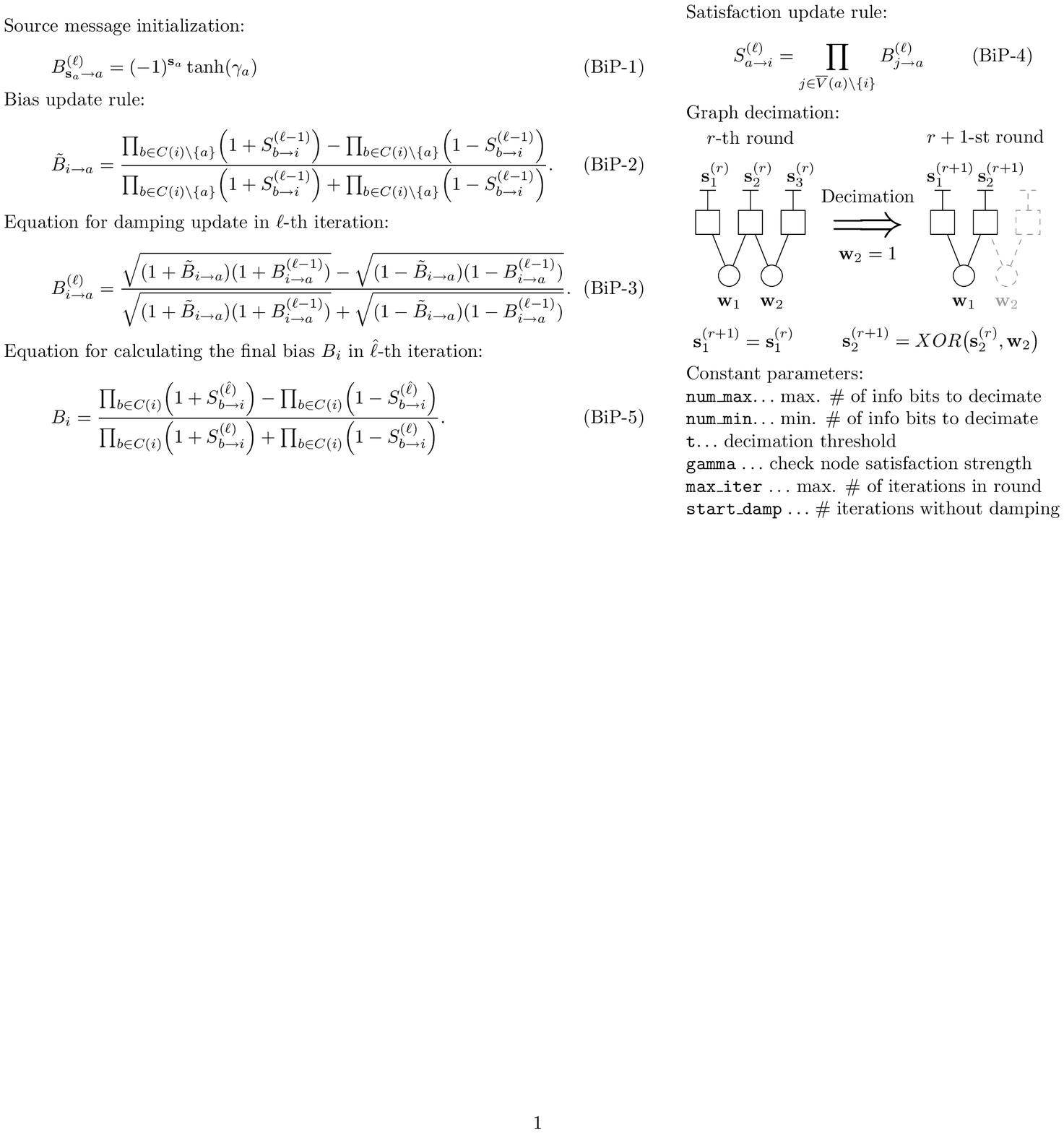}
	\end{minipage}
	}
	\end{center}
	\caption{Summary of the Bias Propagation algorithm.}
	\label{fig:bip-summary}
\end{figure}
The BiP is an iterative message-passing algorithm that performs bitwise MAP estimation of the optimal vector $\vecWs$ for a given source sequence $\vect{s}$. This is done in rounds. In $r$-th round, we use the factor graph $\factorGraph^{(r)}$ and source sequence $\vect{s}^{(r)}$ to find the most probable bits of $\vect{w}_\vect{s}$ to be fixed. The estimation of these bits is done by \pmaxiter~message-passing iterations over the factor graph $\factorGraph^{(r)}$. In $\ell$-th iteration, the bias messages $\bia{\ell}$ and the constant source messages $\bsaa{\ell}$ are sent from information bits and source bits to connected check nodes. Check nodes send satisfaction messages $\sai{\ell}$ to their connected information bits. Finaly, the most probable information bits are fixed and removed from the graph by the decimation process. This results in a new factor graph $\factorGraph^{(r+1)}$ and a source sequence $\vect{s}^{(r+1)}$. We describe the algorithm in a condensed form in Figure~\ref{fig:bip-summary}. Later in this section, we derive the update rules.

Given the original factor graph $\factorGraph^{(1)}=\factorGraph$ and the initial source sequence $\vect{s}^{(1)}=\vect{s}$, we start the message-passing process by setting $\sai{0}=\bsaa{0} ,\; \forall a\in\checks$. The source messages are calculated using equation (BiP-1) and stay constant within one round. The parameter $\gamma_a$ expresses the strength of the check node $a$. After \pmaxiter~message-passing iterations using equations (BiP-2)--(BiP-4), we calculate the final bias $B_i$ for each information bit $i$ using equation (BiP-5). The bias $B_i$ expresses the difference of marginal probabilities $P(\vect{w}_i=0|\vect{s};\vecGamma)-P(\vect{w}_i=1|\vect{s};\vecGamma)$. We fix the most biased information bit to $\vect{w}_i=0$ if $B_i>0$ and $\vect{w}_i=1$ otherwise. Based on the maximal $|B_i|$, we fix \pnummin~or \pnummax~ information bits in each round. The decimation step removes all fixed information bits from the factor graph along with all adjacent edges and checks with degree zero. This results in a new factor graph $\factorGraph^{(2)}$. The new source sequence $\vect{s}^{(2)}$ is obtained from $\vect{s}^{(1)}$ by XOR-ing it with all bits that were connected to the fixed information bits (see Figure~\ref{fig:bip-summary}). The satisfaction messages $\sai{0}$ in the second round are initialized with the value of $\sai{\hat{\ell}}$ messages from the last iteration in the previous round. The decimation step only removes some edges and vertices from the factor graph, but preserves the values of messages associated with each edge. We repeat the described procedure (one round) until we fix all information bits from the original factor graph and output the sequence of information bits as $\vecWs$.

\subsection{Derivation of the BiP algorithm} \label{subsec:bip-derivation}
We now carry out the derivations for an arbitrary profile $\varrho$ assuming that we know the vector $\vecGamma =(\gamma_1,\ldots,\gamma_n)$, $\gamma_i\geq0$ $\forall i=1,\ldots,n$, and reformulate the weighted binary quantization problem as bitwise Maximum A Posteriori (MAP) estimation.

First, we factorize the probability distribution (\ref{eq:bip-probability-distribution}). For $\vect{c}\in\code$, we can find $\vect{w}\in\{0,1\}^m$, such that $\vect{c}=\matG\vect{w}$. Thus, we can write
\begin{align}
P(\vect{w}|\vect{s};\vecGamma) = \prod_{a\in\checks} \psi_a(\vect{w}_{V(a)}, \vect{s}_a), \label{eq:bip-probability-distribution-2}
\end{align}
where $\psi_a\bigl(\vect{w}_{\infobits(a)}, \vect{s}_a\bigr) = e^{\gamma_a}$ if $\vect{s}_a = \sum_{i\in\infobits(a)} \vect{w}_i$ and $\psi_a\bigl(\vect{w}_{\infobits(a)}, \vect{s}_a\bigr) = e^{-\gamma_a}$ otherwise. We will use the sum-product (Belief Propagation) algorithm \cite{Ksch01} to calculate marginal probabilities for each information bit and find the assignment for each information bit in the form
\begin{align}
\vect{\hat{w}}_i =
\arg\max_{\vect{w}_i\in\{0,1\}} P(\vect{w}_i|\vect{s}) =
\arg\max_{\vect{w}_i\in\{0,1\}} \sum_{\sim \vect{w}_i} P(\vect{w}|\vect{s}) =
\arg\max_{\vect{w}_i\in\{0,1\}} \sum_{\sim \vect{w}_i} \prod_{a\in\checks} \psi_a\bigl(\vect{w}_{\infobits(a)}, \vect{s}_a\bigr). \label{bip-sum-product-arg-max}
\end{align}
Here, the sum over all information variables without the $i$-th one is shortened as $\sum_{\sim \vect{w}_i}$.

To calculate (\ref{bip-sum-product-arg-max}) using the sum-product algorithm efficiently, we simplify the original update equations (see \cite{Ksch01} for the original update rules). For our problem, all messages $M_{i\rightarrow a}^{(\ell)}$ and $M_{a\rightarrow i}^{(\ell)}$ passed in $\ell$-th iteration in the original sum-product algorithm are two-component vectors. Using the original update equations, we define
\begin{align}
R_{i\rightarrow a}^{(\ell)} =
\frac{M_{i\rightarrow a}^{(\ell)}(1)}{M_{i\rightarrow a}^{(\ell)}(0)} =
\frac{\prod_{b\in C(i) \setminus \{a\}} M_{b \rightarrow i}^{(\ell-1)}(1)}{\prod_{b\in C(i) \setminus \{a\}} M_{b \rightarrow i}^{(\ell-1)}(0)} =
\prod_{b\in C(i) \setminus \{a\}} R_{b\rightarrow i}^{(\ell-1)} \label{eq:bip-ria}
\end{align}
and for messages $\vect{M}_{a\rightarrow i}$
\begin{align}
R_{a\rightarrow i}^{(\ell)} =
\frac{M_{a\rightarrow i}^{(\ell)}(1)}{M_{a\rightarrow i}^{(\ell)}(0)} =
\frac{\sum_{\vect{w}_{V(a)\setminus \{i\}}} \Bigl[ \psi_a(1,\vect{w}_{V(a)\setminus \{i\}},\vect{s}_a) \prod_{j\in V(a) \setminus \{i\}} M_{j \rightarrow a}^{(\ell)}(\vect{w}_j) \Bigr]}{\sum_{\vect{w}_{V(a)\setminus \{i\}}} \Bigl[ \psi_a(0,\vect{w}_{V(a)\setminus \{i\}},\vect{s}_a) \prod_{j\in V(a) \setminus \{i\}} M_{j \rightarrow a}^{(\ell)}(\vect{w}_j) \Bigr]}. \label{eq:bip-rai}
\end{align}
To calculate the last ratio, we will use the definition of function $\psi_a$ and partition the set $\vect{w}_{V(a)\setminus \{i\}} = W_{even} \cup W_{odd}$, where $W_{even} = \bigl\{ \vect{x}\in\{0,1\}^{|V(a)|-1} \bigl| \textrm{ \# of 1 in } \vect{x} \textrm{ is even} \bigr\}$ and $W_{odd} = \bigl\{ \vect{x}\in\{0,1\}^{|V(a)|-1} \bigl| \textrm{ \# of 1 in } \vect{x} \textrm{ is odd} \bigr\}$.

We now assume that $\vect{s}_a=0$ and later remove this assumption. In this special case, we have $\psi_a(0,\vect{w}_{V(a)\setminus \{i\}},\vect{s}_a)=e^{\gamma_a}$ for all vectors from $W_{even}$ and $\psi_a(0,\vect{w}_{V(a)\setminus \{i\}},\vect{s}_a)=e^{-\gamma_a}$ for all vectors from $W_{odd}$. Conversely, $\psi_a(1,\vect{w}_{V(a)\setminus \{i\}},\vect{s}_a)=e^{-\gamma_a}$ for $W_{even}$ and $\psi_a(1,\vect{w}_{V(a)\setminus \{i\}},\vect{s}_a)=e^{\gamma_a}$ for $W_{odd}$. We can substitute and write $ R_{a\rightarrow i}^{(\ell)} = \frac{e^{-\gamma_a} A + e^{\gamma_a} B}{e^{\gamma_a} A + e^{-\gamma_a} B}$, where
$ A = \sum_{W_{even}} \prod_{j\in V(a) \setminus \{i\}} M_{j \rightarrow a}^{(\ell)}(\vect{w}_j)$, and $ B = \sum_{W_{odd}} \prod_{j\in V(a) \setminus \{i\}} M_{j \rightarrow a}^{(\ell)}(\vect{w}_j)$.
We can express both sums ($A$ and $B$) in terms of the ratios $R_{i\rightarrow a}^{(\ell)}$ by dividing them by the constant factor $\prod_{j\in V(a) \setminus \{i\}} M_{j \rightarrow a}^{(\ell)}(0)$
\begin{align*}
\hat{A} & =
\frac{A}{\prod_{j\in V(a) \setminus \{i\}} M_{j \rightarrow a}^{(\ell)}(0)} =
\sum_{W_{even}} \prod_{j\in V(a) \setminus \{i\}} \frac{M_{j \rightarrow a}^{(\ell)}(\vect{w}_j)}{M_{j \rightarrow a}^{(\ell)}(0)} =
\sum_{W_{even}} \prod_{j\in V(a) \setminus \{i\}} \Bigl( R_{j \rightarrow a}^{(\ell)} \Bigr)^{\vect{w}_j} =\\
& = \frac{1}{2} \Bigl[ \underbrace{\prod_{j\in V(a) \setminus \{i\}} \bigl( 1+R_{j \rightarrow a}^{(\ell)} \bigr)}_{=C} + \underbrace{\prod_{j\in V(a) \setminus \{i\}} \bigl(1-R_{j \rightarrow a}^{(\ell)} \bigr)}_{=D} \Bigr] =
\frac{1}{2}(C+D).
\end{align*}
Similarly, for $B$, $\hat{B} = \frac{1}{2}(C-D)$.
Finally, we obtain
\begin{align}
R_{a\rightarrow i}^{(\ell)} & =
\frac{e^{-\gamma_a} \hat{A} + e^{\gamma_a} \hat{B}}{e^{\gamma_a} \hat{A} + e^{-\gamma_a} \hat{B}} =
\frac{1-\frac{e^{\gamma_a}-e^{-\gamma_a}}{e^{\gamma_a}+e^{-\gamma_a}} \frac{D}{C}}{1+\frac{e^{\gamma_a}-e^{-\gamma_a}}{e^{\gamma_a}+e^{-\gamma_a}} \frac{D}{C}} =
\frac{1-S_{a\rightarrow i}^{(\ell)}}{1+S_{a\rightarrow i}^{(\ell)}}, \label{eq:rai}
\end{align}
where we used the substitution
\begin{align}
	S_{a\rightarrow i}^{(\ell)} = (-1)^{\vect{s}_a}\Bigl(\frac{e^{\gamma_a}-e^{-\gamma_a}}{e^{\gamma_a}+e^{-\gamma_a}} \Bigr) \prod_{j\in V(a) \setminus \{i\}} \frac{1-R_{j \rightarrow a}^{(\ell)}}{1+R_{j \rightarrow a}^{(\ell)}}. \label{eq:sai}
\end{align}
It is easy to see that the case where $\vect{s}_a=1$ can be captured by the given substitution. Using the substitution $ B_{i\rightarrow a}^{(\ell)} = (1-R_{i \rightarrow a}^{(\ell)})/(1+R_{i \rightarrow a}^{(\ell)}) $, we can completely rewrite equations (\ref{eq:bip-ria}) and (\ref{eq:bip-rai}) solely in terms of $B_{i\rightarrow a}^{(\ell)}$ and $S_{a\rightarrow i}^{(\ell)}$ obtaining thus the final message-passing rules. From (\ref{eq:rai}),
\begin{align*}
B_{i\rightarrow a}^{(\ell)} & =
\frac{1-R_{i \rightarrow a}^{(\ell)}}{1+R_{i \rightarrow a}^{(\ell)}} =
\frac{1-\prod_{b\in C(i) \setminus \{a\}} R_{b\rightarrow i}^{(\ell-1)}}{1+\prod_{b\in C(i) \setminus \{a\}} R_{b\rightarrow i}^{(\ell-1)}} =
\frac{1-\prod_{b\in C(i) \setminus \{a\}} \frac{1-S_{b\rightarrow i}^{(\ell-1)}}{1+S_{b\rightarrow i}^{(\ell-1)}}}{1+\prod_{b\in C(i) \setminus \{a\}} \frac{1-S_{b\rightarrow i}^{(\ell-1)}}{1+S_{b\rightarrow i}^{(\ell-1)}}} = \\
& = \frac{\prod_{b\in C(i) \setminus \{a\}} \bigl[ 1+S_{b\rightarrow i}^{(\ell-1)}\bigr] -  \prod_{b\in C(i) \setminus \{a\}} \bigl[1-S_{b\rightarrow i}^{(\ell-1)}\bigr]}{\prod_{b\in C(i) \setminus \{a\}} \bigl[1+S_{b\rightarrow i}^{(\ell-1)}\bigr] + \prod_{b\in C(i) \setminus \{a\}} \bigl[1-S_{b\rightarrow i}^{(\ell-1)}\bigr] }
\end{align*}
and from (\ref{eq:sai})
\begin{align*}
S_{a\rightarrow i}^{(\ell)} =  \prod_{j\in \allbits(a) \setminus \{i\}} B_{j\rightarrow a}^{(\ell)},
\end{align*}
where the source message $B_{\vect{s}_a\rightarrow a}^{(\ell)}$ is defined as $ B_{\vect{s}_a\rightarrow a}^{(\ell)} = (-1)^{\vect{s}_a}\tanh(\gamma_a). $
After \pmaxiter$\,$ iterations, we compute the final bias $B_i$ using the last satisfaction messages $\sai{\hat{\ell}}$
{\small \begin{align*}
B_{i} & =
\frac{P(0)-P(1)}{P(0)+P(1)} =
\frac{1-\frac{P(1)}{P(0)}}{1+\frac{P(1)}{P(0)}} =
\frac{1-\prod_{b\in C(i)} R^{(\hat{\ell})}_{b \rightarrow i}}{1+\prod_{b\in C(i)} R^{(\hat{\ell})}_{b \rightarrow i}} =
\frac{\prod_{b\in C(i)} \bigl[ 1+\sbi{\hat{\ell}}\bigr] -  \prod_{b\in C(i)} \bigl[1-\sbi{\hat{\ell}}\bigr]}{\prod_{b\in C(i)} \bigl[1+\sbi{\hat{\ell}}\bigr] + \prod_{b\in C(i)} \bigl[1-\sbi{\hat{\ell}}\bigr] }.
\end{align*}}
Thus, we just obtained equations (BiP-1), (BiP-2), (BiP-4), and (BiP-5) from Figure~\ref{fig:bip-summary}.
%------------------------------------------------------------------
\subsection{Dealing with cycles in the factor graph} \label{subsec:bip-formal-damping}
The sum-product algorithm is exact (gives exact results) when the underlying graph is a tree. However, many researchers reported good results even for graphs with cycles. In principle, short cycles cause the messages to oscilate. The oscillations can be suppressed using a procedure called ``damping''. A similar approach was introduced in the context of statistical mechanics by Pretti \cite{Pre05}.

Using equation (\ref{eq:bip-ria}), we can write
\begin{align}
\ln R_{i\rightarrow a}^{(\ell)} =
\ln \prod_{b\in C(i) \setminus \{a\}} R_{b\rightarrow i}^{(\ell-1)} =
\sum_{b\in C(i) \setminus \{a\}} \ln R_{b\rightarrow i}^{(\ell-1)}. \label{eq:bip-damping-ln-1}
\end{align}
In other words, the update rule (BiP-2) is a sum of logarithmic terms. Thus, we can use the arithmetic mean in this representation to calculate the output message in the $\ell$-th iteration by averaging the input messages from iterations $\ell-1$ and $\ell-2$. This "low-pass temporal filter" will prevent large changes to output messages. Using (\ref{eq:bip-damping-ln-1}), we can write the output ratio after applying the damping procedure as
$
\ln R_{i\rightarrow a}^{(\ell)} =
\frac{1}{2}\Bigl( \sum_{b\in C(i) \setminus \{a\}} \ln R_{b\rightarrow i}^{(\ell-1)} + \sum_{b\in C(i) \setminus \{a\}} \ln R_{b\rightarrow i}^{(\ell-2)} \Bigr), \label{eq:bip-damping-ln}
$
and hence
\begin{align}
R_{i\rightarrow a}^{(\ell)} = \Bigl( \prod_{b\in C(i) \setminus \{a\}} R_{b\rightarrow i}^{(\ell-1)} \cdot \prod_{b\in C(i) \setminus \{a\}} R_{b\rightarrow i}^{(\ell-2)} \Bigr)^{\frac{1}{2}}. \label{eq:bip-ria-damp}
\end{align}
Using $\bia{\ell}=(1-R_{i\rightarrow a}^{(\ell)})/(1+R_{i\rightarrow a}^{(\ell)})$ and equations (\ref{eq:bip-ria-damp}) and (\ref{eq:rai}), we obtain equation (BiP-3) that is used for damping in the BiP algorithm. To obtain the message $\bia{\ell}$ in practice, we calculate the temporary bias message $\hat{B}_{i\rightarrow a}$ using equation (BiP-2). Finaly, $\bia{\ell}$ is obtained from messages $\hat{B}_{i\rightarrow a}$ and $\bia{\ell-1}$ using equation (BiP-3).

%------------------------------------------------------------------
\section{Convergence of the BiP algorithm}
While deriving the BiP algorithm, we tacitly assumed that for some information bits the magnitude of the bias at the end of each round is high, $|B_i|\approx 1$. Such bits have a high tendency to be fixed to some value without generating too much distortion. To obtain a small final distortion, this condition should be fulfiled in each round for some bits. From practical experiments, we know that there exist good degree distributions satisfying this condition, while other distributions, such as regular distributions, do not build up sufficient bias magnitudes in each round and thus produce very high distortion. This observation was also made in \cite{Wai05}.

We say that the BiP algorithm converges in $r$-th round if there exists at least one information bit $i\in\infobits$ that fulfills $|B_i|>\pthreshold$ for some constant threshold $\pthreshold$. Moreover, we say that the BiP algorithm converges if it converges in all its rounds. A part of our future research is to find a condition for the class of degree distributions for which the BiP algorithm converges. This condition could be used for construction of degree distributions optimized for the BiP algorithm in a manner similar to density evolution in analysis of LDPC codes for channel coding \cite{Mct07}.

Murayama \cite{Mur04} developed an approach for lossy source coding based on a modified Belief Propagation algorithm and provided results for regular LDGM codes with a fixed check degree 2. This algorithm was based on a standard approach used in channel coding: run the BP algorithm over the factor graph of an LDGM code and threshold all bits based on the final LLRs. However, this approach cannot be used for a general LDGM code. The key point here is the degree of all check nodes. This idea is strongly connected with the BiP algorithm. Due to sufficient number of check nodes with degree 2, the BiP algorithm converges in its first round. The same observation can be made in other rounds.

We can think of the BiP algorithm as a generalized approach described by Murayama. In terms of the BiP algorithm, he only uses one round and decimates all information bits at once. The BiP algorithm utilizes more rounds via decimation and thus it can use more general degree distributions. By using irregular degree distributions, we can achieve near-optimal rate-distortion performance. Good LDGM codes can be obtained from degree distributions optimized for channel coding, especially for the BSC channel. This choice, which was also suggested in \cite{Wai05}, was motivated by the work of Martinian et al. \cite{Mar04} who pointed out a tight connection between capacity achieving codes for the BEC channel and their dual codes used for binary quantization in the erasure case.

\begin{figure}[t]
	{\begin{minipage}{15.3cm} \small
	Rate: 0.37\\
	$\rho(x)=0.2710 x +0.2258 x^{2} +0.1890 x^{5} +0.0614 x^{6} +0.2528 x^{13}$\\
	$\lambda(x)=0.9522 x^{9} +0.0478 x^{10}$ \vspace{0.2cm}

	Rate: 0.5\\
	$\rho(x)=0.1787 x +0.1762 x^{2} +0.1028 x^{5} +0.1147 x^{6} +0.0122 x^{12} +0.0479 x^{13} +0.1159 x^{14} +$\\
	\hspace*{1.01cm} $+0.2516 x^{39}$\\
	$\lambda(x)=0.9988 x^{9} +0.0012 x^{10}$ \vspace{0.2cm}

	Rate: 0.65\\
	$\rho(x)=0.2454 x +0.1921 x^{2} +0.1357 x^{5} +0.0838 x^{6} +0.1116 x^{12} +0.0029 x^{14} +0.0222 x^{15} +$\\
	\hspace*{1.01cm} $+0.0742 x^{28} +0.1321 x^{32}$\\
	$\lambda(x)=0.4987 x^{5} +0.5013 x^{6}$ \vspace{0.2cm}

	Rate: 0.75\\
	$\rho(x)=0.2912 x +0.1892 x^{2} +0.0408 x^{4} +0.0873 x^{5} +0.0074 x^{6} +0.1126 x^{7} +0.0926 x^{15} +$\\
	\hspace*{1.01cm} $+0.0187 x^{20} +0.1241 x^{32} +0.0361 x^{39}$\\
	$\lambda(x)=0.8016 x^{4} +0.1984 x^{5}$
	\end{minipage}}

	\caption{List of good degree distributions used for generating the results.}
	\label{fig:degree-distribution-1}
\end{figure}

%------------------------------------------------------------------
\section{Results}
We implemented the BiP algorithm in C++, where the messages were represented using single precision numbers. The update equations were manually optimized using Intel's SSE1 extension so that the cache memory was used in an optimal way. All results presented in this work were obtained using an Intel Core2 X6800 2.93GHz CPU machine with 2GB RAM. We ran the machine on Linux in 64 bit mode. All C++ code was optimized to 64 bit mode and compiled using Intel C++ 9.0 compiler. The speed of this algorithm (throughput) was measured while both CPU cores were utilized. We ran the same algorithm on both cores, resulting in a 1.7--1.8 times higher throughput.

In Figure~\ref{fig:degree-distribution-1}, we present degree distributions from edge perspective that were used for generating all results. Some distributions were obtained from the LdpcOpt site (\url{http://lthcwww.epfl.ch/research/ldpcopt/}) and optimized for the BSC channel. %To evaluate the convergence condition of the degree distribution for each round, we solved equations (\ref{eq:decimation-unnorm-1})--(\ref{eq:decimation-unnorm-3}) for some finite $n$ using simple recursion. For evaluating the performance of a given degree distribution when used with the BiP algorithm, we define the \emph{delay of degree distribution} as the percentage of info bits that need to be removed from the graph so that the BiP algorithm will converge in the next round ($\max_{i\in\infobits} |B_i|>$\pthreshold).

%\input{fig_dd_conv_analysis}
%In Figure~\ref{fig:convergence-analysis}, we plot the convergence condition for the (4,8) regular distribution and an irregular degree distribution for rate $R=0.5$ from Figure~\ref{fig:degree-distribution-1}. Codes based on regular degree distributions have very large delays. In the convergence graph, the delay as obtained from the intersection of 1 and the convergence curve. From practical experiments of running the BiP algorithm, we estimate the delay from quantizing 30 random source sequences ($\gamma=1$) and obtain 31.8\%. This value is close to the value what we can estimate from the graph. Other listed irregular degree distributions have similar convergence graph as the degree distribution of rate $R=0.5$. For a uniform profile, we selected the parameter $\gamma$ ($\vecGamma=(\gamma,\ldots,\gamma)$) so that the BiP algorithm converged in the first round. We took $\gamma=0.8, 1.07, 1.3, 1.5$ for rates $R=0.37, 0.5, 0.65, 0.75$, respectively.

To compare our results with the work of Wainwright et al. \cite{Wai05}, we implemented their algorithm while using similar optimization approaches. In the case of an irregular code with rate $R=0.5$ and length $n=5000$, we achieved throughput $1013$ bits/sec. The BiP algorithm produced the same distortion with throughput $10415$ bits/sec. The BiP algorithm is a special case of their algorithm for $\wsou=\winfo=0$.

\begin{figure}[t]
	\begin{center} \hspace*{-0.5cm}
		{\begin{minipage}{6.7cm}
			\vspace*{0.3cm}
			%X axis
			\psfrag{0.3}[tc][tc]{\scriptsize $0.3$} \psfrag{0.4}[tc][tc]{\scriptsize $0.4$}
			\psfrag{0.5}[tc][tc]{\scriptsize $0.5$} \psfrag{0.6}[tc][tc]{\scriptsize $0.6$}
			\psfrag{0.7}[tc][tc]{\scriptsize $0.7$} \psfrag{0.8}[tc][tc]{\scriptsize $0.8$}
			\psfrag{0.9}[tc][tc]{\scriptsize $0.9$} \psfrag{1}[tc][tc]{\scriptsize $1$}
			%Y axis
			\psfrag{0}[Br][Br]{\scriptsize $0$} \psfrag{0.02}[Br][Br]{\scriptsize $0.02$}
			\psfrag{0.04}[Br][Br]{\scriptsize $0.04$} \psfrag{0.06}[Br][Br]{\scriptsize $0.06$}
			\psfrag{0.08}[Br][Br]{\scriptsize $0.08$} \psfrag{0.1}[Br][Br]{\scriptsize $0.1$}
			\psfrag{0.12}[Br][Br]{\scriptsize $0.12$} \psfrag{0.14}[Br][Br]{\scriptsize $0.14$}
			\psfrag{0.16}[Br][Br]{\scriptsize $0.16$} \psfrag{0.18}[Br][Br]{\scriptsize $0.18$}
			\psfrag{0.2}[Br][Br]{\scriptsize $0.2$}
		
			\psfrag{xlabel}[tc]{\small Rate $R=\frac{m}{n}$}
			\psfrag{ylabel}[bc][tc]{\small Distortion $D$}
			\psfrag{bound}[cc]{}
			\psfrag{bound2}[cc]{}
			\psfrag{title}[Bc][tc]{\small Rate-distortion performance}
		
			\includegraphics[width=6.4cm, clip]{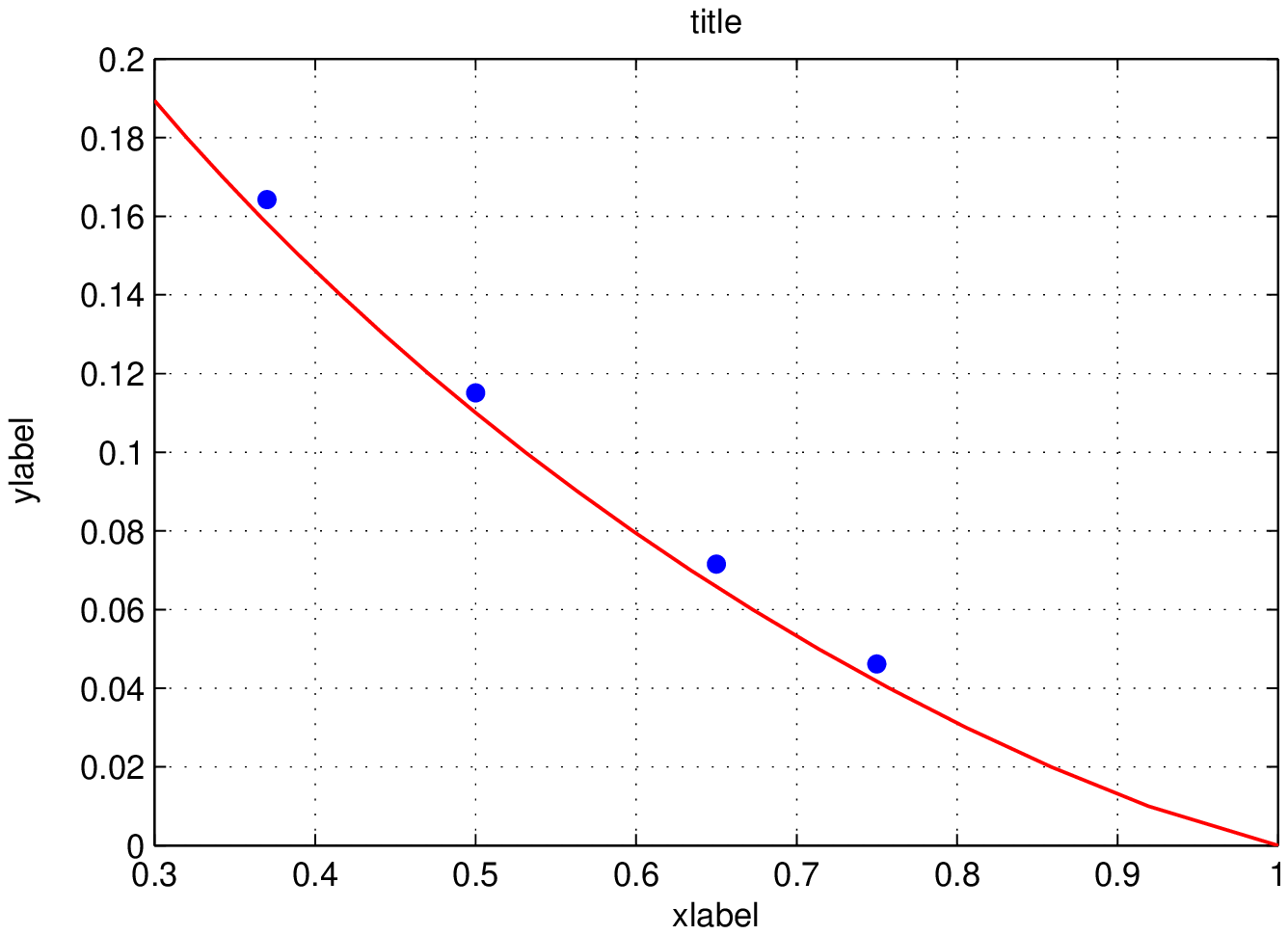}
			\put(-190,120){\small(a)}
		\end{minipage}}
		{\begin{minipage}{8.1cm}
			%X axis
			\psfrag{X3}[tc][tc]{\scriptsize $10^3$} \psfrag{X4}[tc][tc]{\scriptsize $10^4$} \psfrag{X5}[tc][tc]{\scriptsize $10^5$}
			%Y axis
			\psfrag{0.112}[Br][Br]{\scriptsize 0.112} \psfrag{0.114}[Br][Br]{\scriptsize 0.114}
			\psfrag{0.116}[Br][Br]{\scriptsize 0.116} \psfrag{0.118}[Br][Br]{\scriptsize 0.118}
			\psfrag{0.12}[Br][Br]{\scriptsize 0.12}   \psfrag{0.122}[Br][Br]{\scriptsize 0.122}
			\psfrag{0.124}[Br][Br]{\scriptsize 0.124} \psfrag{0.126}[Br][Br]{\scriptsize 0.126}
			\psfrag{0.11}[Br][Br]{\scriptsize 0.11}   \psfrag{ 0.12}[Br][Br]{\scriptsize 0.12}

			\psfrag{2000}[Br][Br]{\scriptsize 2000} \psfrag{4000}[Br][Br]{\scriptsize 4000}
			\psfrag{6000}[Br][Br]{\scriptsize 6000} \psfrag{8000}[Br][Br]{\scriptsize 8000}
			\psfrag{10000}[Br][Br]{\scriptsize 10000} \psfrag{12000}[Br][Br]{\scriptsize 12000}

			\psfrag{xlabel}[tc]{\small Code length $n$}
			\psfrag{bound}[cl]{\scriptsize Rate-distortion bound}

			\psfrag{Throughput}[Bc][tc]{\small Throughput (bits/sec.)}
			\psfrag{Distortion}[Bc][tc]{\small Distortion}

			\includegraphics[viewport=40 0 510 270, width=8.3cm, clip]{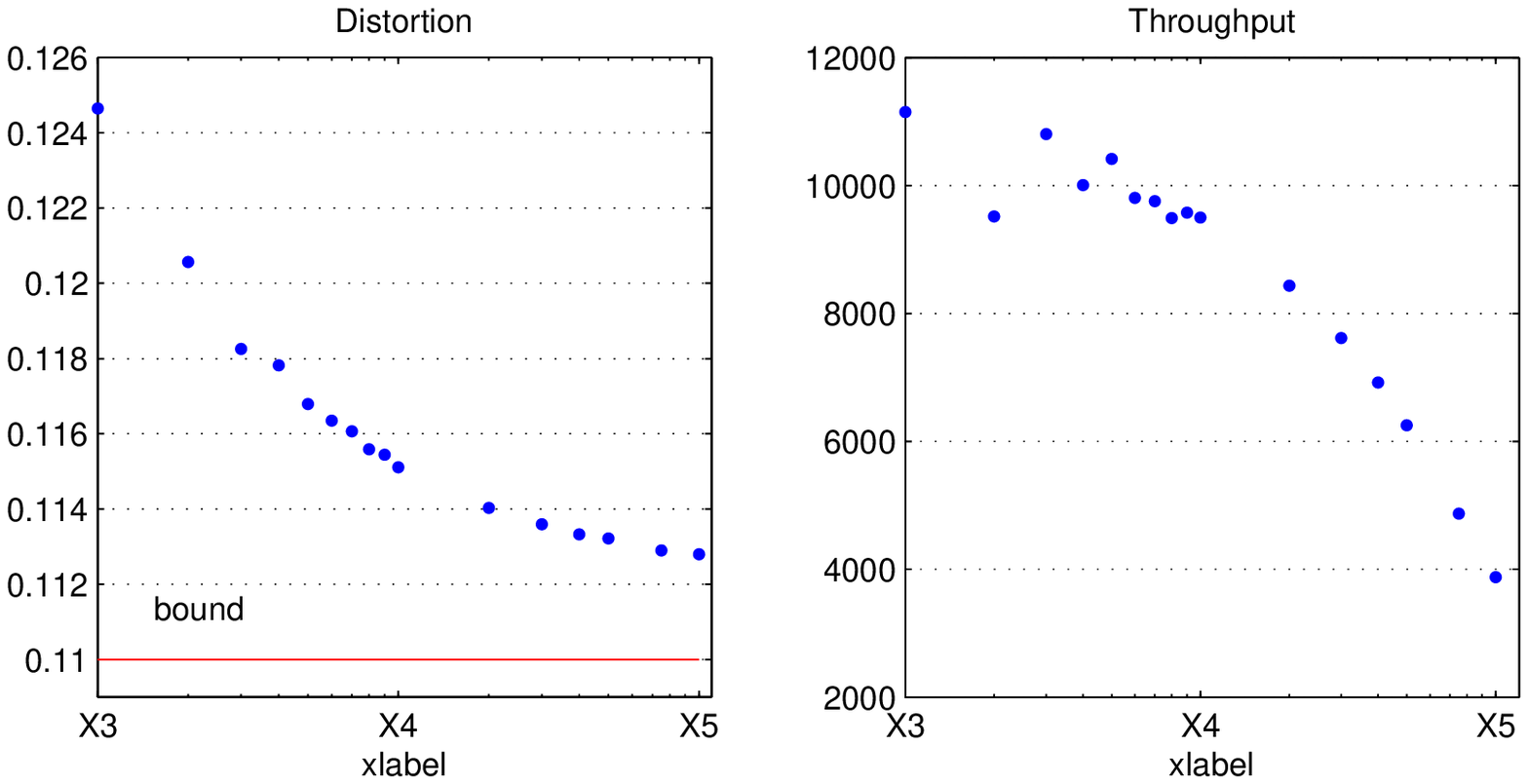}
			\put(-240,120){\small(b)}
		\end{minipage}}
	\end{center}

	\caption{(a) rate-distortion performance plot for selected irregular codes of length $n=10^4$. (b) distortion and throughput plots for various code lengths for irregular code with rate $R=0.5$. In all graphs, each point represents an average over 100 trials.}
	\label{fig:rate-distortion-graph}
\end{figure}

Figure~\ref{fig:rate-distortion-graph} contains simulation results for various rates and various code lengths for a uniform profile $\varrho$. All results were generated using the following parameter values: $\pthreshold=0.8$, $\pmaxiter=25$, $\pstartdamp=10$, $\pnummax=0.01\times m$, $\pnummin=0.001\times m$. We set $\vecGamma=(\gamma,\ldots,\gamma)$, where $\gamma=0.8$, $\gamma=1.07$, $\gamma=1.3$, $\gamma=1.5$ for rates $R=0.37$ to $R=0.75$, respectively.

%Up to now, we presented the results for uniform binary quantization. As discussed in Section~\ref{sec:bip-weighted-case}, we can use the BiP algorithm with a non-constant $\gamma$ and hence perform weighted binary quantization. Here, we should be carefull in setting $\gamma_a$ to each check node $a\in\checks$. From convergence analysis, we have the requirement on the variance of $\ddens{b}_\vect{s}$. In practice, we can change $\gamma_a$, but the variance of $\ddens{b}_\vect{s}$ should be equal to the original variance with constant $\gamma$.

We now present the results for a linear profile of weights (which is important for applications in steganography \cite{Fri07Spie}), and perform weighted binary quantization using the BiP algorithm. For a source sequence $\vect{s}$ of length $n$, we define the linear profile $\varrho=(\varrho_1,\ldots,\varrho_n)$ as $\varrho_i=2(n-i)/n$. %The distortion is now defined as $D=\sum_{i=1}^n\varrho_i|\vect{s}_i-\vect{\hat{s}}_i|$, where $\vect{\hat{s}}$ is the reconstructed sequence.
From \cite{Fri07Spie} (Appendix A), the rate-distortion bound in this weighted case will be achieved if each bit is flipped with probability $p_a(1)=e^{-\zeta \varrho_{a}}/(1+e^{-\zeta \varrho_{a}})$ for each source bit $a\in\checks$.
% JF: Navrhuji pouzit zpet \zeta, nebot \lambda znamena v tomto paperu predevsim degree distribution. Paklize to akceptujes, nezapomen zmenit ten symbol taky v patricnych figures.
% TF: OK mas pravdu.
% TF: zmenil jsem tvar toho p_a(1). Mel by to dle mych informaci byt podil e^{...}/(1+e^{...}) a ne jen e^{...}.

% JF: Mohl bys zmenit "theoretical bound" na "p_a(1)" ve Figure 5? Nemeli bychom p_a(1) rikat bound. Je to ta pravdepodobnost, se kterou by se melo flipovat, aby se dosahla rate-distortion.
% TF: hotovo

% For a given $m$, $n$, and profile $\varrho$, we find $\zeta$ using binary search. equation (\ref{eq:minim-weighted-case-pi})), we solve equation (\ref{eq:minim-weighted-case-sum}) and find the corresponding parameter $\zeta$ for given rate. This can be done easily using binary search. When we know $\zeta$, we can obtain the lower bound on $p_a(1)$ for each source bit $a\in\checks$ (substitute $\zeta$ to (\ref{eq:minim-weighted-case-pi})).
To find optimal values of $\gamma_a$, we use the following iterative approach. Start with $\gamma_a^{(0)}=\gamma$, where $\gamma$ is taken from the uniform weight case. Find an estimate of $p_a(1)$ (denote it $\hat{p}_a(1)$) by quantizing $k$ random source sequences. We calculate the estimate as an arithmetic average and finally use a moving average filter to smooth the resulting sequence. Comparing the estimate $\hat{p}_a(1)$ with $p_a(1)$, we finally set $\gamma_a^{1} = \gamma_a^{0}+c\bigl(\hat{p}_a(1)-p_a(1)-(\hat{p}-p)\bigr)$, where $\hat{p}$ and $p$ is the arithmetic average of $\hat{p}_a(1)$ and $p_a(1)$, respectively, and $c$ is a constant (in practice, we use $c=3$). 

\begin{figure}
	\psfrag{title1}[Bc][Bc]{\small Values of $\gamma_a$}
	\psfrag{title2}[Bc][Bc]{\small Flipping probabilities, $\hat{p}_a(1)$, $p_a(1)$}

	%X axis
	\psfrag{X0}[tr][tr]{\small 0} \psfrag{1000}[tc][tc]{\small 1000} \psfrag{2000}[tc][tc]{\small 2000}
	\psfrag{3000}[tc][tc]{\small 3000} \psfrag{4000}[tc][tc]{\small 4000} \psfrag{5000}[tc][tc]{\small 5000}
	\psfrag{6000}[tc][tc]{\small 6000} \psfrag{7000}[tc][tc]{\small 7000} \psfrag{8000}[tc][tc]{\small 8000}
	\psfrag{9000}[tc][tc]{\small 9000} \psfrag{10000}[tr][tr]{\small 10000}
	%Y axis
	\psfrag{0}[cr][cr]{\small 0} \psfrag{1}[cr][cr]{\small 1}\psfrag{2}[cr][cr]{\small 2}\psfrag{3}[cr][cr]{\small 3}
	\psfrag{0.2}[cr][cr]{\small 0.2}\psfrag{0.4}[cr][cr]{\small 0.4}\psfrag{0.6}[cr][cr]{\small 0.6}\psfrag{0.8}[cr][cr]{\small 0.8}

	\psfrag{xlabel}[tc][Bc]{\small $a$ - source bit id}
	\psfrag{xlabel1}[tc][Bc]{\small $a$ - source bit id, $n=10000$, $f(x)=0.0792x^3-0.0841x^2-0.7925x+1.3378$}
	\psfrag{ylabel1}[Bc]{}
	\psfrag{ylabel2}[Bc]{\small \hspace*{0.5cm} $\convcond(\ddens{b}_\vect{s}, \rho(t), \lambda(t))$}

	\psfrag{polynom}[cl][cl]{\footnotesize $\gamma_a=f\Bigl(\frac{a-n/2}{0.2886n}\Bigr)$}
	\psfrag{estimated probabilityxxxxxxxxxx}{\footnotesize estimated probability $\hat{p}_a(1)$}
	\psfrag{bound}{\footnotesize $p_a(1)$}

	{\includegraphics[width=15.5cm, clip]{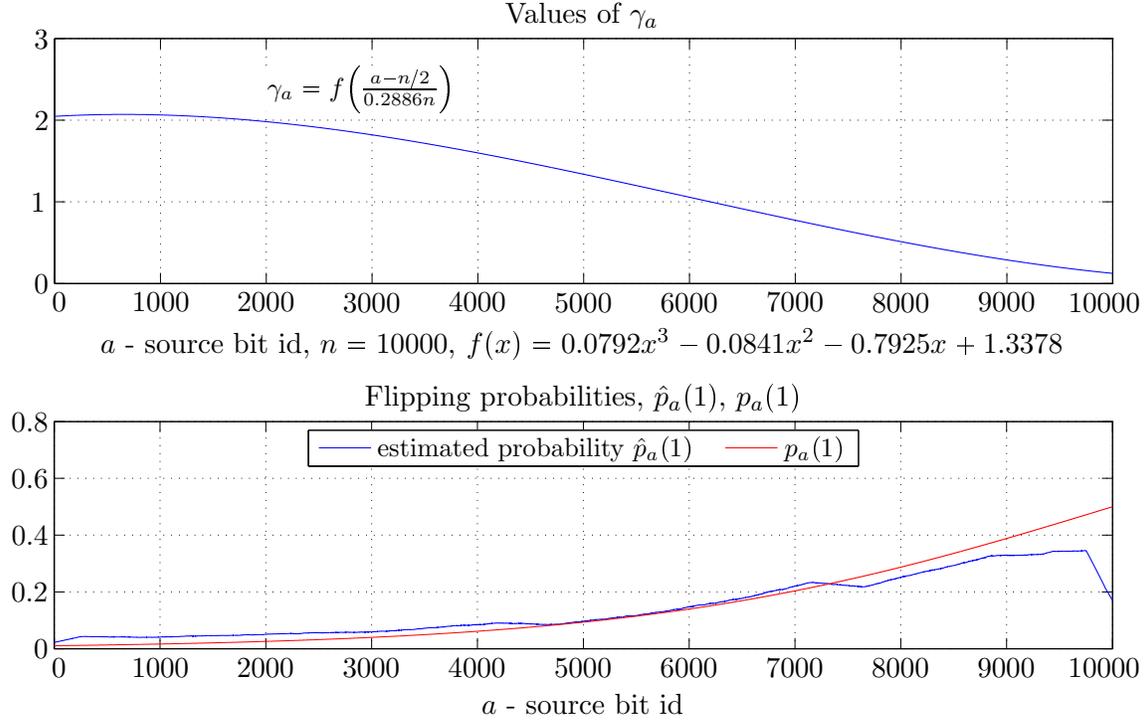}}

	\caption{Flipping probabilities for linear profile $\varrho$, $R=0.5$, $\zeta=4.544$. Values of $\gamma_a$ were obtained using an iterative approach for $n=10000$ and interpolated by a cubic polynomial.}
	\label{fig:results-weighted-case-gamma}
\end{figure}

\begin{figure}\begin{center}
	{\begin{minipage}{7.25cm}
		\psfrag{title}[Bc][tc]{\small Rate-distortion performance}
	
		%X axis
		\psfrag{0}[Br][Br]{\scriptsize 0} \psfrag{0.02}[cr][cr]{\scriptsize 0.02} \psfrag{0.04}[cr][cr]{\scriptsize 0.04}
		\psfrag{0.06}[cr][cr]{\scriptsize 0.06}  \psfrag{0.08}[cr][cr]{\scriptsize 0.08}  \psfrag{0.1}[cr][cr]{\scriptsize 0.1}
		\psfrag{0.12}[cr][cr]{\scriptsize 0.12} \psfrag{0.14}[cr][cr]{\scriptsize 0.14} \psfrag{0.16}[cr][cr]{\scriptsize 0.16}
		\psfrag{0.18}[cr][cr]{\scriptsize 0.18}
		%Y axis
		\psfrag{0.3}[tl][tl]{\scriptsize 0.3} \psfrag{0.4}[tc][tc]{\scriptsize 0.4} \psfrag{0.5}[tc][tc]{\scriptsize 0.5}
		\psfrag{0.6}[tc][tc]{\scriptsize 0.6} \psfrag{0.7}[tc][tc]{\scriptsize 0.7} \psfrag{0.8}[tc][tc]{\scriptsize 0.8}
		\psfrag{0.9}[tc][tc]{\scriptsize 0.9} \psfrag{1}[tc][tc]{\scriptsize 1}
	
		\psfrag{Rate}[tc]{\small Rate $R=\frac{m}{n}$}
		\psfrag{Distortion}[bc][tc]{\small Weighted distortion}
		\psfrag{ylabel1}[Bc]{}
		\psfrag{ylabel2}[Bc]{\small \hspace*{0.5cm} $\convcond(\ddens{b}_\vect{s}, \rho(t), \lambda(t))$}
	
		\psfrag{boundxxxxxxxxxxxxxxxx}{\scriptsize theoretical bound}
		\psfrag{wbip}{\scriptsize weighted BiP}
		\psfrag{bip}{\scriptsize ordinary BiP}

		{\includegraphics[viewport=-20 -20 430 320, width=7.2cm, clip]{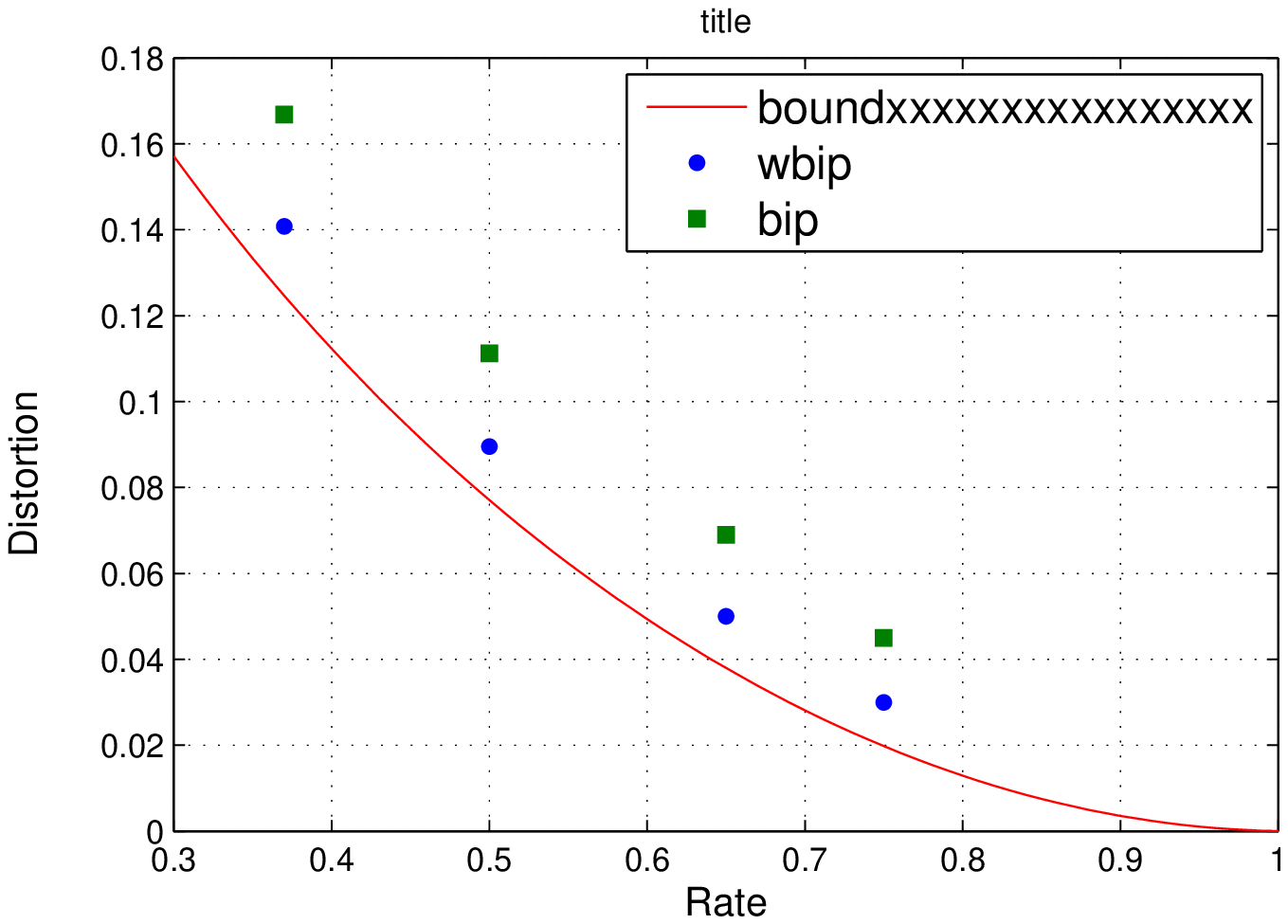}}
		\put(-204,138){\small(a)}
	\end{minipage}}
	{\begin{minipage}{8.1cm}
		%X axis
		\psfrag{X3}[tc][tc]{\scriptsize $10^3$} \psfrag{X4}[tc][tc]{\scriptsize $10^4$} \psfrag{X5}[tc][tc]{\scriptsize $10^5$}
		%Y axis
		\psfrag{0.089}[Br][Br]{\scriptsize 0.089}
		\psfrag{0.09}[cr][cr]{\scriptsize 0.09}
		\psfrag{0.091}[cr][cr]{\scriptsize 0.091}
		\psfrag{0.092}[cr][cr]{\scriptsize 0.092}
		\psfrag{0.093}[cr][cr]{\scriptsize 0.093}
		\psfrag{0.094}[cr][cr]{\scriptsize 0.094}
		\psfrag{0.095}[cr][cr]{\scriptsize 0.095}
		\psfrag{0.096}[cr][cr]{\scriptsize 0.096}

		\psfrag{3000}[Br][Br]{\scriptsize 3000} \psfrag{4000}[cr][cr]{\scriptsize }
		\psfrag{5000}[cr][cr]{\scriptsize 5000} \psfrag{6000}[cr][cr]{\scriptsize }
		\psfrag{7000}[cr][cr]{\scriptsize 7000} \psfrag{8000}[cr][cr]{\scriptsize }
		\psfrag{9000}[cr][cr]{\scriptsize 9000} \psfrag{10000}[cr][cr]{\scriptsize }
		\psfrag{11000}[cr][cr]{\scriptsize 11000} \psfrag{12000}[cr][cr]{\scriptsize }
		\psfrag{13000}[cr][cr]{\scriptsize 13000}

		\psfrag{xlabel}[tc]{\small Code length $n$}

		\psfrag{Distortion}[Bc][tc]{\small Weighted distortion}
		\psfrag{Throughput}[Bc][tc]{\small Throughput (bits/sec.)}

		\includegraphics[viewport=30 -4 640 336, width=8.3cm, clip]{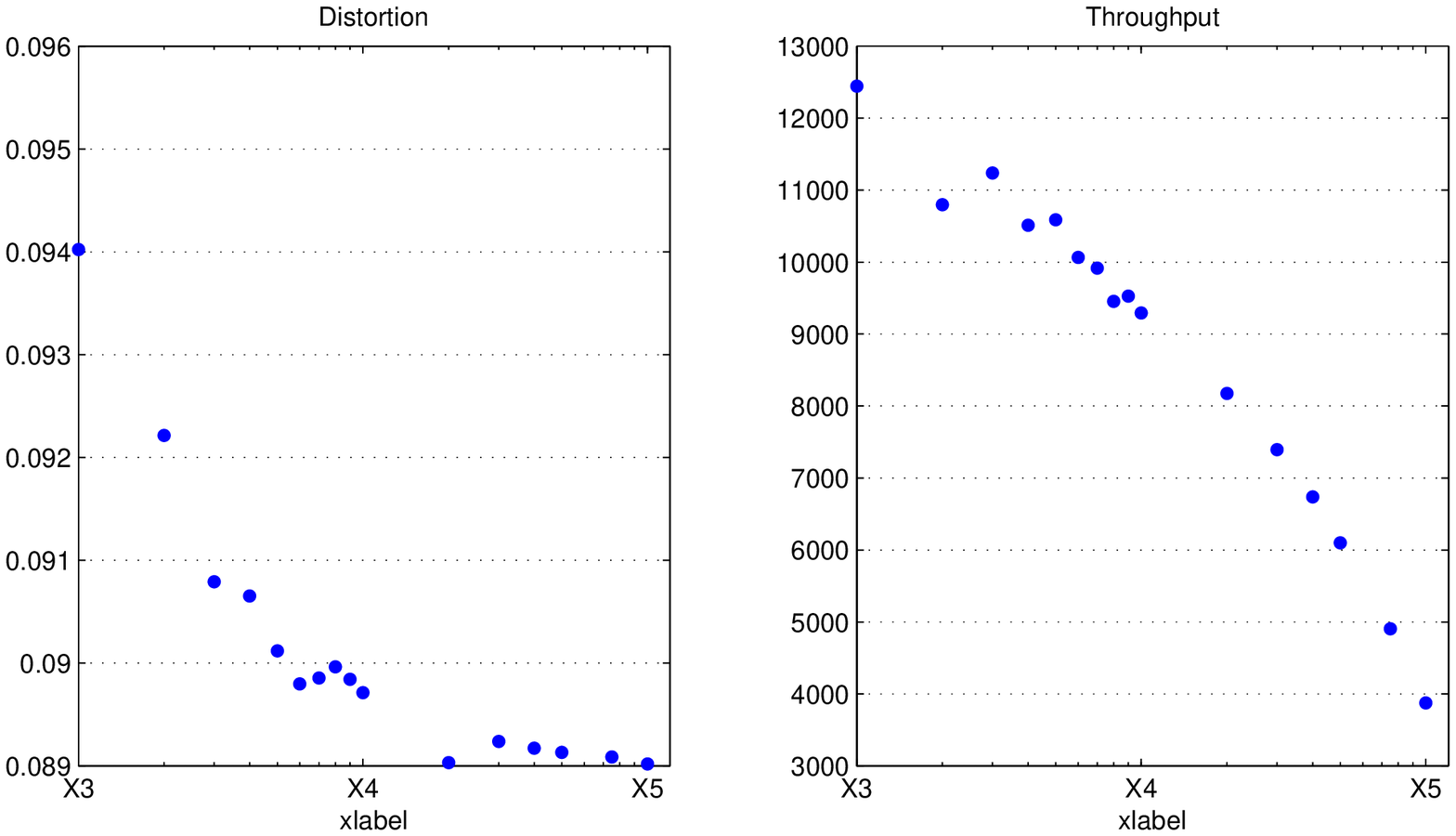}
		\put(-250,130){\small(b)}
	\end{minipage}}

	\caption{(a) overall comparison of weighted vs. non-weighted BiP algorithm for a linear profile $\varrho$, $n=10000$. (b) results from using weighted BiP algorithm for linear profile $\varrho$ using different code lengths, $R=0.5$. %Values of $\gamma_a$ for each check node $a\in\checks$ were obtained using the polynomial fit from Figure~\ref{fig:results-weighted-case-gamma}.
	In all graphs, each point represents an average over 100 trials.}
	\label{fig:results-weighted-case-results}
\end{center}\end{figure}

% This approach uses the difference between the estimate and the lower bound to change the sequence of $\gamma_a$ values. 
Usually, 10 iterations were sufficient to obtain good results. In Figure~\ref{fig:results-weighted-case-gamma}, we present the resulting $\hat{p}_a(1)$ and $\gamma_a$ for rate $R=0.5$. A cubic polynomial was fit through the data points (it uses a centralized variable $x$). In Figure~\ref{fig:results-weighted-case-results} (b), we use this polynomial and show how the BiP depends on the code length. We can see that the throughput is roughly the same as for the non-weighted BiP. In Figure~\ref{fig:results-weighted-case-results} (a), we present the overall comparison of weighted vs. non-weighted BiP algorithm for all degree distributions. Here, we use the ordinary BiP algorithm ($\vecGamma=(\gamma,\ldots,\gamma)$) and measure the distortion using the weighted norm (linear case). Although the degree distributions were taken from Figure \ref{fig:degree-distribution-1} and were thus not optimized for the weighted case, the weighted BiP algorithm still achieves very good results.
% JF: Ta centralized variable x je definovana jako by "a" byla random variable (std(a), mean(a), atd.), ale nikde se nedefinuje jakozto r.v. Je treba to formulovat nejak strucne, ale presne, aby to nebylo confusing.
% TF: hotovo. Zadefinoval jsem ten polynom, jako funkci f(x) v labelu grafu a do grafu dal jeji vypocet.


\section{Conclusions and future work}
We proposed a new algorithm for binary quantization based on the Belief Propagation algorithm with decimation over factor graphs of LDGM codes. We call the algorithm Bias Propagation (BiP). Using the Bias Propagation algorithm, we significantly reduced the complexity of binary quantization using LDGM codes in comparison with \cite{Wai05}. We believe this reduction is new and constitutes an important contribution as it allows us to theoretically study the algorithm in the future. We postpone this problem to our future work. We showed that the algorithm based on pure Belief Propagation has the same rate-distortion performance as the algorithm based on Survey Propagation \cite{Wai05}.%In our analysis, we applied the tools used in density evolution. We derived a necessary condition for the BiP algorithm to converge. This condition describes the form of LDGM codes that can be used with this algorithm.

%We believe that the BiP algorithm is amenable to theoretical analysis necessary convergence condition is strong enough to develop a search procedure for finding optimal degree distributions for binary quantization using LDGM codes in a manner similar to density evolution.

\bibliographystyle{plain}
\bibliography{all_coding,all_stego}

\end{document}